\documentclass[sigconf]{acmart}
\AtBeginDocument{%
  \providecommand\BibTeX{{%
    \normalfont B\kern-0.5em{\scshape i\kern-0.25em b}\kern-0.8em\TeX}}}
\acmConference[ICSE 2024]{46th International Conference on Software Engineering}{April 2024}{Lisbon, Portugal}

\copyrightyear{2024}
\acmYear{2024}
\setcopyright{acmlicensed}\acmConference[ICSE '24]{2024 IEEE/ACM 46th International Conference on Software Engineering}{April 14--20, 2024}{Lisbon, Portugal}
\acmBooktitle{2024 IEEE/ACM 46th International Conference on Software Engineering (ICSE '24), April 14--20, 2024, Lisbon, Portugal}
\acmDOI{10.1145/3597503.3639150}
\acmISBN{979-8-4007-0217-4/24/04}

\usepackage{listings}
\usepackage{tcolorbox}
\usepackage{soul}
\usepackage{xcolor}
\usepackage{multirow}
\usepackage{fancybox}
\usepackage{xspace}
\usepackage{ragged2e}
\usepackage{changepage}
\usepackage{makecell}
\usepackage{setspace}
\usepackage{enumitem}

\usepackage{booktabs, multirow} 
\usepackage{soul}

\usepackage{url}
\usepackage{hyperref}
\hypersetup{
    colorlinks=false,
    linkcolor=blue,
    anchorcolor=blue,
    filecolor=blue,      
    citecolor=blue,
    urlcolor=blue
}

\usepackage{amsmath,amssymb,amsfonts}

\usepackage{textcomp}
\usepackage{lipsum}
\usepackage{graphicx}
\usepackage{subfigure}
\usepackage{balance}
\usepackage[ruled, lined, linesnumbered, commentsnumbered, longend]{algorithm2e}

\usepackage{algorithmic}
\SetCommentSty{mycommfont}
\usepackage{array}

\usepackage{adjustbox}
\usepackage{threeparttable}

\newcommand{\phead}[1]{\vspace{1mm} \noindent {\bf #1}}

\newcommand{\tsmall}{LLMParser$_{T5Small}$\xspace}
\newcommand{\tbase}{LLMParser$_{T5Base}$\xspace}
\newcommand{\llama}{LLMParser$_{LLaMA}$\xspace}
\newcommand{\chatGLM}{LLMParser$_{ChatGLM}$\xspace}
\newcommand{\logllm}{LLMParser\xspace}
\newcommand{\logllms}{LLMParsers\xspace}

\newcommand{\rqboxc}[1]{\begin{tcolorbox}[left=1pt,right=1pt,top=0pt,bottom=0pt,colback=gray!5,colframe=gray!40!black,before skip=5pt,after skip=0pt]#1\end{tcolorbox}}

\definecolor{lightgray}{gray}{0.9}

\newcommand\greybox[1]{%
  \vskip\baselineskip%
  \vspace{-0.8\baselineskip}
  \par\noindent\colorbox{lightgray}{%
    \begin{minipage}{\linewidth}#1\end{minipage}%
  }%
  \vspace{-0.8\baselineskip}
  \vskip\baselineskip%
}

\lstdefinelanguage{XML}
{
  morestring=[b]",
  morestring=[s]{>}{<},
  morecomment=[s]{<?}{?>},
  stringstyle=\color{black},
  identifierstyle=\color{darkblue},
  keywordstyle=\color{cyan},
  morekeywords={xmlns,xsi,schemaLocation}
}

\usepackage{array}
\newcolumntype{$}{>{\global\let\currentrowstyle\relax}}
\newcolumntype{^}{>{\currentrowstyle}}

\newcolumntype{?}{!{\vrule width 1.5pt}}

\begin{document}

\title{\logllm: An Exploratory Study on Using Large Language Models for Log Parsing}

\author{Zeyang Ma}
\affiliation{%
\institution{Software PErformance, Analysis and Reliability (SPEAR) Lab}
 \institution{Concordia University}
 \city{Montreal}
 \state{Quebec}
 \country{Canada}}
\email{m_zeyang@encs.concordia.ca}

\author{An Ran Chen}
\affiliation{%
\institution{Electrical and Computer Engineering Department}
 \institution{University of Alberta}
 \city{Edmonton}
 \state{Alberta}
 \country{Canada}}
\email{anran6@ualberta.ca}

\author{Dong Jae Kim}
\affiliation{%
\institution{Software PErformance, Analysis and Reliability (SPEAR) Lab}
 \institution{Concordia University}
 \city{Montreal}
 \state{Quebec}
 \country{Canada}}
\email{k_dongja@encs.concordia.ca}

\author{Tse-Hsun (Peter) Chen}
\affiliation{%
\institution{Software PErformance, Analysis and Reliability (SPEAR) Lab}
 \institution{Concordia University}
 \city{Montreal}
 \state{Quebec}
 \country{Canada}}
\email{peterc@encs.concordia.ca}

\author{Shaowei Wang}
\affiliation{%
\institution{Department of Computer Science}
 \institution{University of Manitoba}
 \city{Winnipeg}
 \state{Manitoba}
 \country{Canada}}
\email{shaowei@cs.umanitoba.ca}



\begin{abstract}
Logs are important in modern software development with runtime information.
Log parsing is the first step in many log-based analyses, that involve extracting structured information from unstructured log data. 
Traditional log parsers face challenges in accurately parsing logs due to the diversity of log formats, which directly impacts the performance of downstream log-analysis tasks.
In this paper, we explore the potential of using Large Language Models (LLMs) for log parsing and propose \logllm, an LLM-based log parser based on generative LLMs and few-shot tuning. We leverage four LLMs, Flan-T5-small, Flan-T5-base, LLaMA-7B, and ChatGLM-6B in \logllms. 
Our evaluation of 16 open-source systems shows that \logllm achieves statistically significantly higher parsing accuracy than state-of-the-art parsers (a 96\% average parsing accuracy). 
We further conduct a comprehensive empirical analysis on the effect of training size, model size, and pre-training LLM on log parsing accuracy. 
We find that smaller LLMs may be more effective than more complex LLMs; for instance where Flan-T5-base achieves comparable results as LLaMA-7B with a shorter inference time. 
We also find that using LLMs pre-trained using logs from other systems does not always improve parsing accuracy. 
While using pre-trained Flan-T5-base shows an improvement in accuracy, pre-trained LLaMA results in a decrease (decrease by almost 55\% in group accuracy).
In short, our study provides empirical evidence for using LLMs for log parsing and highlights the limitations and future research direction of LLM-based log parsers. 

\end{abstract}
    


\settopmatter{printfolios=true}
\maketitle

\section{Introduction}
\label{sec:introduction}

\begin{figure}
\centering
\includegraphics[width=\columnwidth]{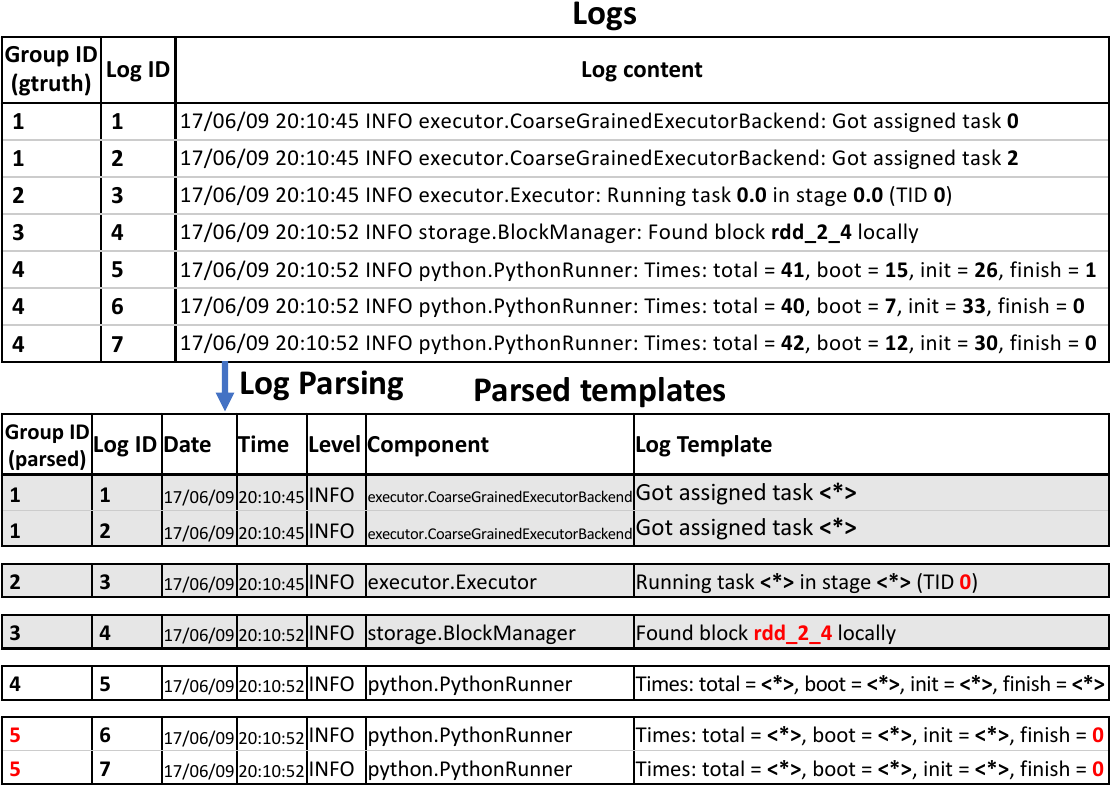}
\vspace{-2.5em}
\caption{An example of log parsing and validating the result from Spark. The incorrectly parsed results are highlighted in red.
}
\label{fig1:definition}
\vspace{-1.5em}
\end{figure}



Software logs provide developers with valuable system runtime information to understand system execution and debugging. 
However, due to the sheer volume and complexity of logs, analyzing logs manually becomes infeasible. To assist with log analysis, researchers have proposed many automated approaches for various tasks such as anomaly detection~\cite{he2016experience,zhang2019robust}, monitoring~\cite{chen2019experience, wang2021would}, and root cause analysis~\cite{yuan2010sherlog,lu2017log,he2021survey}. 
Among all log analysis tasks, log parsing often serves as the first step of log analysis.



The goal of log parsing is to convert raw log data into log templates by identifying and separating static text and variable values in the log messages.
As shown in Figure~\ref{fig1:definition}, logs contain dynamic information such as the timestamp, log level, and log message (which contains static message and dynamic variable values). Log parsing first extracts consistent information among all the logs using regular expression (e.g., timestamps and log level), and then transforms logs into a more structured format (i.e., log template) by identifying variables in the log message~\cite{li2023did,he2017towards}.
For instance, 
the log message in Log 1 from Figure~\ref{fig1:definition} has the log message \texttt{Got assigned task 0} and the log can be parsed to the log template \texttt{Got assigned task <*>} with an identified variable \texttt{0}. 
The variables record system runtime information that can be in various forms (e.g., string, digits, or symbols). Including such dynamic variable values in the logs makes automated log analysis difficult. Hence, log parsing is an essential first step in log analysis, and having low accuracy in log parsing results directly impacts the performance of downstream tasks~\cite{li2023did, Drain, he2017towards}.

Despite the importance of log parsing, effectively parsing logs remains a challenging task. There are many prior research that proposed various log parsers~\cite{SLCT, LogClusterC, dai2020logram, LKE, Drain}. Yet, recent studies~\cite{zhu2019tools,10.1145/3510003.3510101} show that these approaches often fail to identify parameters in logs, which may affect the downstream log analysis tasks. 
Recently, Large Language Models (LLMs) have demonstrated promising results in text-related and code-related tasks, such as code understanding and generation~\cite{leinonen2023code-understanding,white2023chatgpt_code}.
Intuitively, log is composed of both natural language and code-like variables. LLMs' strong ability for language translation can be potentially leveraged for log parsing, which can also be viewed as translating from logs to log templates.  

In this paper, we investigate the potential of using LLMs for log parsing, with a focus on studying the effect of varying LLMs, shot sizes, and pre-training
particularly when working with limited training data. We proposed \logllm, an innovative log parser. \logllms learn from few-shot examples on how to ``translate'' a log into a log template and evaluate \logllm using four text-to-text or text generation LLMs: Flan-T5-small~\cite{flan-t5}, Flan-T5-base~\cite{flan-t5}, LLaMA-7B~\cite{touvron2023llama}, and ChatGLM-6B~\cite{zeng2023glm-130b}. 
We train and evaluate \logllms using a widely-used log benchmark that contains logs data from 16 open-source systems~\cite{he2020loghub,10.1145/3510003.3510101}.
Our evaluation shows that 1) \logllms can achieve an average parsing accuracy (PA) of 0.96, which is higher state-of-the-arts parsers Drain~\cite{Drain}, Logram~\cite{dai2020logram}, and LogPPT~\cite{le2023log} (among them, the highest PA is 0.92). 
2) Few-shot tuning is more effective than in-context learning, where in-context learning only results in an average PA of 0.46. 3) Increasing the number of training examples may not always give better parsing results; data diversity and balance may be more important. 4) More complex LLMs may not always give better results. We find that Flan-T5-base, which only has 240M parameters, can achieve similar results compared to LLaMA which has 7B parameters. 5) LLM pre-trained using logs from other systems may not always help improve PA. We find opposite findings between Flan-T5-base and LLaMA, where LLaMA experiences a decrease in parsing accuracy while Flan-T5-base has an improved parsing result.

We summarize the main contributions of this paper as follows:
\setlist[itemize]{leftmargin=*}
\begin{itemize}
  \item We explore the use of LLMs for log parsing, and propose \logllm, a generative LLM-based approach for log parsing. \logllm achieves a higher parsing accuracy (PA) compared to state-of-the-arts. 
  \item We compare in-context learning and few-shot tuning and found that few-shot tuning achieves a much higher PA (up to 0.96 v.s. 0.46). We also found that few-shot tuning is efficient, which only takes from one to five minutes on an NVIDIA A100 GPU. 
  \item We found that increasing training shot sizes may not always improve PA. Future studies should explore better sampling approaches to improve LLM-based log parsers. 
  \item LLMs with more parameters may not always give better PA. We find that a medium-size LLM (Flan-T5-base) achieves comparable performance compared to LLaMA-7B. Future studies should consider the trade-off between model complexity and accuracy. 
  \item We find that using LLMs pre-trained using logs from other systems may not necessarily improve PA. We saw contradictory results in LLaMA and Flan-T5-base, where the parsing accuracy using LLaMA decreases. Future studies are needed to explore the impact and effectiveness of pre-trained log models on log parsing.  
\end{itemize}

\phead{Paper Organization.} Section~\ref{sec:background} discusses background and related work. Section~\ref{sec:Approach} provides the details of \logllm. Section~\ref{sec:setup} shows experiment setup and our implementation. Section~\ref{sec:evaluation} evaluates \logllm. Section~\ref{sec:Discussion} discusses the implications of our findings. Section~\ref{sec:threat} discusses threats to validity. Section~\ref{sec:conclusion} concludes the paper. 

\phead{Data Availability:} We made our source code and experimental results
publicly available at: \url{https://github.com/zeyang919/LLMParser}

\section{Background and Related Work}
\label{sec:background}
In this section, we discuss the background of 
Large Language Models (LLMs) and how to optimize LLMs on specific tasks. We also discuss related work, existing log parsing approaches, and applications that use LLMs to solve log-related tasks.

\subsection{Background}

\phead{Large Language Models.}
The Large Language Models (LLMs) are mostly developed based on the transformer architecture~\cite{t5, touvron2023llama,zeng2023glm-130b}. 
LLMs have made important advancements in natural language processing (NLP) by providing models that have an extraordinary capacity for understanding language and producing contextually relevant and semantically consistent text.
LLMs are generally pre-trained on a large corpus of text data from diverse sources such as books, articles, websites, and even source code.
Recent studies~\cite{leinonen2023code-understanding,white2023chatgpt_code} have highlighted the capability of LLMs in code recognition, understanding, and code generation. 

As logs consist of both natural language sequences and code-like variables, prior work~\cite{chen2022bert, dai2020logram, he2018characterizing, li2018studying} has leveraged language models to analyze logs. 
However, it remains unclear whether LLMs can be effectively used for log parsing due to the varying pre-training data and model characteristics.
Adopting LLMs for log parsing brings potential advances in the research area. 
First, LLMs are shown to be very powerful on text-related tasks~\cite{leinonen2023code-understanding,white2023chatgpt_code}, which may be able to achieve more accurate log parsing results. 
Second, LLMs are generalizable on unseen data~\cite{9609166,wang2021language,8952475}, which may be leveraged to parse new logs without continuous retraining. 
Nevertheless, there is a need for a comprehensive study on using LLMs for log parsing and what kinds of adaptions are needed for logs. 

\phead{In-context Learning and Fine-Tuning of Large Language Models.}
To adapt LLMs to specific tasks, there are two main strategies: in-context learning and few-shot tuning. In-context learning~\cite{brown2020gpt3,wang2020generalizing} is a method that incorporates task-specific demonstrations directly into the input (i.e., prompt) during LLMs’ inference, guiding the model to generate responses in a desired manner without the need for retraining/changing the model’s parameters. In-context learning relies on the model's ability to generalize from the provided demonstrations to understand and execute the task at hand. On the other hand, fine-tuning~\cite{dodge2020fine,radford2018improving,gao-etal-2021-making} involves re-training the pre-trained LLM on a dataset tailored to the specific task, allowing the model to adjust its internal parameters and better align its outputs with the desired outcomes. In particular, few-shot tuning~\cite{liu2022few,mosbach2023few} is a fine-tuning method that enables LLMs to generalize from limited examples, which may facilitate the extraction of relevant information across diverse log formats, including log variables, log structure, and semantic patterns.


While in-context learning provides quick adaptability, it has several drawbacks. First, processing prompts with several demonstrations every time the model parses logs can contribute to further computational costs. 
In-context learning prompted with few-shot demonstrations requires the model to process both the target instance and all the demonstrations during each inference, leading to an increased inference time.
Second, the context size of the model’s inputs limits the number of demonstrations that can be used. For example, performing in-context learning with four prompts on Flan-T5-Base~\cite{flan-t5} exceeds its context size of 512 tokens. This limitation poses a challenge for LLMs to learn from a larger number of demonstrations and improve their performance. Finally, selecting effective demonstrations is also crucial for improving the performance of in-context learning, as it is sensitive to the format and order of the prompts~\cite{dong2022survey,min2022rethinking,lu2021fantastically}.

In contrast, few-shot tuning does not demand continuous in-context demonstrations for every inference, which can speed up inference time. 
Moreover, using fine-tuning, we can provide more diverse log examples to train the model as the tuning is already performed during training. 
Prior studies~\cite{brown2020gpt3,liu2022few,gao-etal-2021-making} have also demonstrated that few-shot tuning offers better accuracy at lower computational costs.
Furthermore, since few-shot tuning only involves a small number of data samples, the entire fine-tuning process can be fast (e.g., only a few minutes for our approach). As a result, there is no significant time overhead incurred due to fine-tuning.
Therefore, in this paper, we utilize few-shot tuning to integrate domain-specific knowledge into LLMs. 

\subsection{Related Work}
We discuss related work along two directions: log parsing and using LLM for other log-related tasks. 

\phead{Log Parsing.}
To support log parsing for large volumes of logs, researchers have proposed many automated log parsing techniques. 
Existing log parsers primarily use three approaches: frequent pattern mining, log clustering, and parsing trees.
(1) Frequent pattern mining identifies static text and variables by counting the number of times a pattern or sequence recurs in the logs (e.g., SLCT~\cite{SLCT},
LogCluster~\cite{LogClusterC}, Logram~\cite{dai2020logram}).
(2) Log clustering groups logs using clustering algorithms, thereby categorizing logs into different groups (e.g., LKE~\cite{LKE}, LogSig~\cite{LogSig}, and LenMa~\cite{lenma}).
(3) Parsing tree builds a parse tree with fixed depth to guide the log group search process (e.g., Drain~\cite{Drain}).
These studies aim to strike a balance between optimizing accuracy and the size of pre-learned data for log parsing tasks. While the accuracy of log parsing has shown improvement over time, recent studies~\cite{zhu2019tools,10.1145/3510003.3510101} reveal that traditional algorithm-based log parsing tools primarily emphasize log clustering.
Although these approaches achieve high grouping accuracy, they often fail to accurately identify all the variables in logs~\cite{10.1145/3510003.3510101}. Therefore, this limitation may hinder downstream log analysis tasks, such as missing some anomalies recorded by unrecognized variables during log anomaly detection~\cite{li2023did}.


The recent rise of LLMs has brought new possibilities for improving log parsing. 
Le and Zhang~\cite{le2023log} proposed LogPPT, which is a log parser based on a masked language model (RoBERTa~\cite{Roberta}). LogPPT improved the accuracy of log parsing compared to traditional algorithm-based log parsers. 
LogPPT converts the log parsing task into a token classification task to classify whether a token in the log is variable or static. However, this process requires more manual effort in labelling every token in a log on whether or not a token is a static text. 
Le and Zhang~\cite{le2023evaluation} further evaluated using ChatGPT~\cite{brown2020gpt3} to parse logs. 
Their study shows that ChatGPT can parse the logs but the accuracy is worse than LogPPT. However, due to the closed-source nature of ChatGPT, the monetary cost can be high, the LLM is not fine-tunable, and the stability of the LLM is out of the control of the developers. More importantly, logs often contain sensitive data that cannot be sent to third parties. 

In this paper, we investigate the application of text generation and text2text generation LLMs to tackle the log parsing task. The recent advancements and ongoing research in LLMs, particularly in text2text and text generation, have significantly improved their text understanding and processing capabilities~\cite{brown2020gpt3,touvron2023llama,chia2023instructeval}. 
Intuitively, log parsing is similar to language translation, where a log is translated into a log template. 
This allows us to eliminate the process of splitting logs and manually labeling individual tokens, and the parsed log template is directly obtained.
We leverage four open-source LLMs (Flan-T5-Base~\cite{flan-t5}, Flan-T5-Small~\cite{flan-t5}, LLaMA-7B~\cite{touvron2023llama}, and ChatGLM-6B~\cite{zeng2023glm-130b}) to generate the log template by inputting the prompt and explore the performance of the LLMs compared with the state-of-the-art approaches.
Furthermore, we study the limitations of LLM-based log parsers and explore the potential of using pre-trained LLM-based log parsers.


\phead{Using LLMs for Other Log-related Tasks.}
The recent advances in LLMs have shown success in both natural language processing and code generation~\cite{rahaman2023chatgpt,leinonen2023code-understanding,white2023chatgpt_code}. 
Since logs are semi-structured text composed of natural language with some code elements, researchers have adopted LLMs for log-related tasks~\cite{liu2023scalable,chen2023empowering,lee2021lanobert}. 
Some studies~\cite{liu2023scalable,lee2021lanobert} used LLMs for log anomaly detection. 
Lee et al.~\cite{lee2021lanobert} proposed LAnoBERT which is an anomaly log detector.
LAnoBERT masked the specific word in the log and then used BERT~\cite{devlin2018bert} to predict the masked word and calculate the predictive probability of the origin masked word. When there are large differences between the actual and the predicted words, the respective log is identified as an anomaly.
Liu et al.~\cite{liu2023scalable} conducted a case study on logs from Huawei Cloud and found that the effectiveness of the
anomaly detected by ChatGPT was partially consistent with that of the on-call engineers, which suggests that LLMs could potentially reduce manual verification efforts. 
Chen et al.~\cite{chen2023empowering} introduced RCACopilot, an on-call system integrated with OpenAI's GPT models for automating root cause analysis of cloud incidents.


\section{Approach}
\label{sec:Approach}

\begin{figure}
\centering 
\includegraphics[width=1\columnwidth]{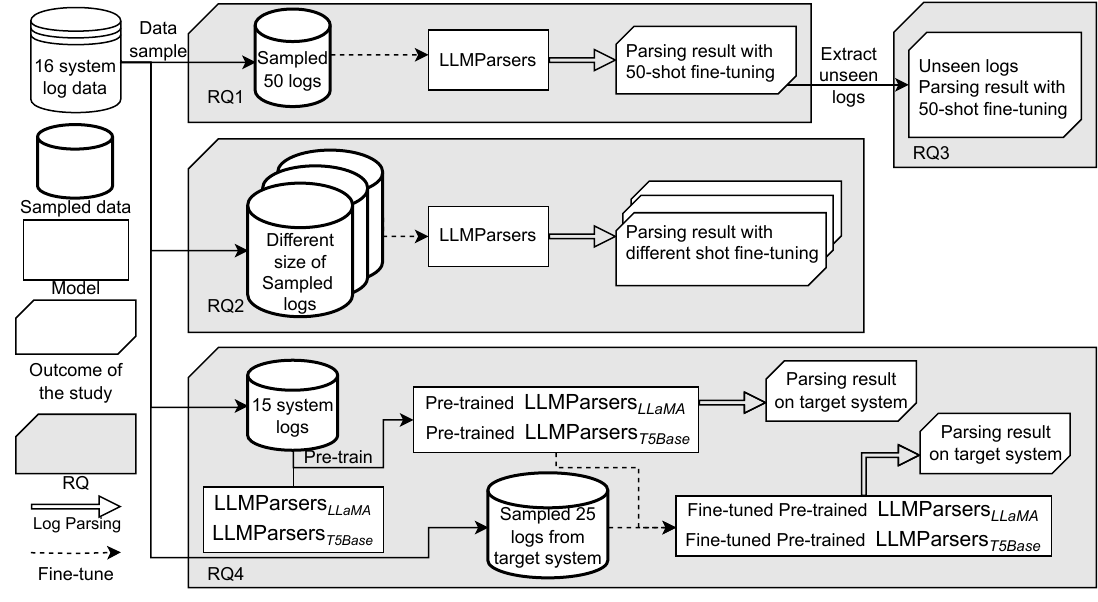} 
\caption{An overview of the evaluation of \logllms. 
}
\label{fig2:overall}
\end{figure}

In this section, we introduce \logllms, which adopts LLMs for log parsing. \logllms tackle log parsing as a text2text and text generation task using few-shot fine-tuning. 
Figure~\ref{fig2:overall} illustrates our overall evaluation architecture of \logllm. 
In RQ1, we sample 50 logs from each system using our sampling algorithm for fine-tuning \logllms, and evaluate the log parsing result. In RQ2, we conduct a sensitivity analysis on the number of shots used for fine-tuning. 
In RQ3, we evaluate the effectiveness of \logllms on unseen log templates. Finally, in RQ4, evaluate the log parsing accuracy of using a pre-trained \logllm using 15 other systems. 
Below, we present our few-shot sampling algorithm, LLM selection, prompt selection, and fine-tuning process. 

\subsection{Sampling Few-Shot Data}
LLMs require data samples to learn how to process different requests, a task often accomplished by providing a few training examples~\cite{le2023log,le2023evaluation}. 
Similar to most other log parsers (e.g., Logram, Drain, and LogPPT), \logllm is an offline parser. Log parsers typically need to scan all available logs to identify patterns and abstract dynamic values, a process that is inherently offline. Access to all necessary logs is a prerequisite for this procedure, enabling the application of various techniques like clustering and log sampling. Hence, we can apply our clustering and log sampling techniques to sample a small number of logs and their associated log templates to train the LLMs. 
We prioritize the sampling of more commonly-seen logs while ensuring data diversity. 
Prior studies~\cite{8717641,10.1145/2897937.2897995} also suggest that increasing the diversity of training data is effective in improving the understanding and generalization ability of deep learning models. 
Therefore, we proposed a data sampling algorithm to sample logs with high frequency and variety to increase the coverage and diversity of the training dataset.

\setlength{\textfloatsep}{0pt}
\begin{algorithm}
\setstretch{0.75}
\SetNoFillComment
\small
    \SetKwFunction{isOddNumber}{isOddNumber}
    \SetKw{KwRet}{return}
    \SetKw{Return}{return}
    \SetKwInOut{KwIn}{Input}
    \SetKwInOut{KwOut}{Output}
    \SetKwFunction{extractcontent}{extract-content}
    \SetKwFunction{replacenumber}{replace-numbers-with-0}
    \SetKwFunction{clustering}{mean\_shift\_clustering}
    \SetKwFunction{reaplceoriginal}{replace-logs-with-original}
    \SetKwFunction{sortbyclustersize}{sort-clusters}
    \SetKwFunction{sorted}{sorted}

    \SetKwFunction{sortbygroupsize}{sort-groups}
    \SetKwFunction{setunsampled}{set-unsampled}
    \SetKwFunction{sampleonelog}{sample\_one\_log}
    \SetKwFunction{grouplogs}{group-logs}
    \SetKwFunction{processRawLogs}{process\_raw\_logs}
    \SetKwFunction{findtemplate}{find\_template}

    \SetKwFunction{getlargestunsampledgroup}{get-largest-unsampled}
    \SetKwFunction{labeltemplate}{label-template}
    \KwIn{$D$: Dataset containing raw logs}
    \KwIn{$N$: Number of logs to sample}
    \KwOut{$D_{train}$: Dataset with $N$ sampled logs and log templates}
    \tcc{\textbf{Extract timestamps and replace numbers}}
    $processed\_logs \leftarrow$     \processRawLogs{$D$};

    
    $log\_clusters \leftarrow$\clustering($processed\_logs$)\;
    $D_{train} \gets \varnothing$;     

    \ForEach{$cluster$ in \sorted($log\_clusters$, ``descend'')}{
        $log$ $\leftarrow$ \sampleonelog{$cluster$}\;
        
        $D_{train}.$\textbf{add} ($log$, \findtemplate{$log$})\;
        \If{length of $D_{train}=N$}{
            \textbf{break}\;
        }
    }


    \KwRet{$D_{train}$}
\caption{Few-shot Log Sampling
}
\label{alg:sample}
\end{algorithm}

Algorithm~\ref{alg:sample} demonstrates our log sampling process. The sampling process does not require any pre-labelled data and is unsupervised.  
Firstly, similar to other log parsers~\cite{9252062,le2023log}, we process the raw logs to separate and remove the information generated by logging frameworks, such as dates and timestamps, and extract the log messages. 
We further process the logs by using regular expressions to replace all the numbers in the log with a unified character to minimize the effect of the dynamically generated values on the next step. 
Secondly, we apply the Mean Shift~\cite{400568} clustering algorithm to cluster 
the processed logs. 
We choose Mean Shift because, unlike K-means, it does not need the user to specify the number of clusters in advance. However, other clustering algorithms can also be used. 
Thirdly, we sort the generated clusters in descending order based on the cluster size. 
Finally, we sample one log from every cluster and repeat the process until we reach the desired number of samples. 
By sampling and iterating from the largest cluster, we consider both diversity and coverage (i.e., possibly covering more logs) in the sampling process.  
We label the log templates for all the sampled logs to guide the LLM on how to parse logs.  


\subsection{\logllm: Using LLMs for Parsing Logs }
\phead{Parsing Logs Using Text2text and Text Generation LLMs.}
When using text2text or text generation LLMs, we can parse a log by simply giving them a log message. The LLMs ``translate'' the log into a log template. 
Compared to LogPPT~\cite{le2023log}, which uses LLMs for token classification, log parsing using text2text and text generation LLMs eliminates the log splitting and output conversion process. The LLMs directly use the logs for input and output, which fully leverages LLMs' abilities and makes the parsing result more intuitive and easier to diagnose parsing issues. 


We evaluate and compare four LLMs on their log parsing accuracy: Flan-T5-small~\cite{flan-t5}, Flan-T5-base~\cite{flan-t5}, LLaMA-7B~\cite{touvron2023llama} and ChatGLM-6B~\cite{zeng2023glm-130b}. Table~\ref{tab:llms} shows the architecture of the LLMs and the parameter size. 
The LLMs cover both text2text (Flan-T5) and text generation (LLaMa and ChatGLM), vary in size (range from 80M to 7B parameters), and are pre-trained using different architectures. 
Our goal is to explore the difference in log parsing accuracy among LLMs with different architectures and parameter sizes. 
Flan-T5-small and Flan-T5-base are the instruction fine-tuned versions of  T5~\cite{t5} with different parameter sizes. Prior research~\cite{TheFlanC28:online,flan-t5} showed that Flan-T5 converges faster than T5 and achieves outstanding results on fine-tuned tasks. 
LLaMA-7B is a publicly released state-of-the-art foundational large language model by Meta, which offers smaller and more efficient models for researchers on multiple NLP tasks.
ChatGLM-6B was jointly built by Tsinghua University and Zhipu AI Company~\cite{zeng2023glm-130b} that enable widespread access to researchers for question answering and information processing.
Additionally, they are open-source models, which allows for easy in-depth analysis and fine-tuning processes, as well as the replication of our study in future research.



\begin{table}
\caption{Information on the Large Language Models that we used for log parsing.
\vspace{-0.4cm}
}\label{tab:llms}
\scalebox{0.78}{
\setlength{\tabcolsep}{0.25cm}{
\begin{tabular}{lllrr}\toprule
LLM Name &Architecture&Pre-training Objective &Parameter Size \\
\midrule
Flan-T5-small & Encoder-decoder &Text2text Generation &80M \\
Flan-T5-base & Encoder-decoder &Text2text Generation &240M \\
LLaMA-7B &Causal Decoder&Text Generation &7B \\
ChatGLM-6B &Prefix Decoder&Text Generation &6B \\
\bottomrule
\end{tabular}}}
\end{table}

\phead{Prompt Templates for Log Parsing.}
Prompts are user-provided inputs, such as queries, instructions, or questions, that guide LLMs and instruct their behaviour for specific tasks. 
Specifically, prompts are incorporated into the model's embedding layer to guide its decision-making process. 
Prompting involves priming an LLM by using prompts to demonstrate examples of the downstream task.
Previous studies~\cite{gao-etal-2021-making,zhou2022large,white2023prompt} have demonstrated that the quality of input prompts plays a crucial role in the performance of LLMs, influencing the generated output quality. 

A prompt template is often used to generate prompts in a consistent and reproducible way. As an example, in the case of log parsing, one possible prompt template can be formulated as ``The log [X] belongs to the log template [Y].'', where the objective is to generate the corresponding log template for the input log [X], with [Y] representing the answer. 
In this paper, we investigate the effectiveness of LLMs in log parsing using hard prompts, which 
are fixed natural language instructions. We use hard prompts~\cite{wen2023hard} to minimize the variability caused by the prompts, and focus our study on examining the impact of different LLMs and varying training sizes on parsing performance. Below, we discuss the prompts that we used in \logllms. 

In the design of its prompt, T5 leverages the colon symbol ``{\sf :}'' to separate instructions and input data.
Since our task shares similarities with the machine translation task, which involves transforming input data to output data, we leverage T5's default prompt structure (i.e., ``instruction + input type + output type:") to build our \tsmall and \tbase prompt template as:
\greybox{\textit{``Parse the raw log to log template: `\{Raw log\}'.''}\\
\textit{``\{Log template\}''}
}
\noindent We feed the log and log template pairs as examples during training. When parsing logs, we only give instructions about the raw log. 

LLaMA-7B and ChatGLM-6B are two large text generation models that have been extensively studied and fine-tuned for various tasks~\cite{fine-tuned_chatglm6b, zhang2023llama_adapter, anand2023gpt4all}.
One notable optimized version of LLaMA is Alpaca~\cite{alpaca}, which is highly regarded for its performance.
The prompt template used by Alpaca has been widely adopted, consistently delivering excellent performance.
In this paper, we use Alpaca's prompt template (defined below) to train and generate output for the log parsing task using LLaMA-7B and ChatGLM-6B models. 
\greybox{\textit{
``Below is an instruction that describes a task, paired with an input that provides further context.
        Write a response that appropriately completes the request.
        \\
        \\
        \#\#\# Instruction:
        Parse the input log to the log template.
        \\
        \#\#\# Input:
        `\{Raw log\}'
        \\
        \#\#\# Response:
        `\{Log Template\}'
        }}
\noindent We give the task description, instruction, input, and response to the LLMs during training. When parsing, we only give the description, instruction, and input, and ask the LLMs to generate the response. 

\phead{Applying Few-shot Tuning on the LLMs.} When the training dataset is the same, the fine-tuning efficiency of a model is directly related to its parameter size, as larger models with more parameters require more computational resources and a longer time to converge~\cite{church2021emerging}.
Due to the small parameter size of the Flan-T5-small and Flan-T5-base, fine-tuning can be efficiently done without additional optimization mechanisms. 
For the larger models, LLaMA-7B and ChatGLM-6B, we used LoRA~\cite{hu2021lora} to accelerate the fine-tuning process. 
LoRA is an efficient parameter fine-tuning technique, which only trains a very small portion of {\em ADDITIONAL} parameters while freezing the original parameters of the model. LoRA uses low-rank parameterization and focuses on the most important layers of models, reducing both computational and memory requirements. 
Consequently, the model converges faster with minimal impact on performance, enabling faster fine-tuning for various tasks, which achieves comparable speed to fine-tuning Flan-t5-base (even with more than 25 times more parameters). Our fine-tuning process takes from several seconds to less than five minutes.

\section{Experiment Setup and Implementation}
\label{sec:setup}
In this section, we discuss our experiment setup to answer our research questions and the implementation details. 

\phead{Studied Dataset.}
We conduct our experiment on the log parsing benchmark provided by He et al.~\cite{he2020loghub}. This benchmark contains logs from 16 open-source systems and is widely used to evaluate and compare the accuracy of log parsers~\cite{dai2020logram,Drain,8489912}. Each system includes 2,000 log messages along with their respective log templates and parameters (the ground truth for evaluating log parsers).
The studied systems included in the dataset range from various domains, such as distributed systems, operating systems, mobile systems, supercomputers, server applications, and standalone software. 
However, recent studies~\cite{dai2020logram,10.1145/3510003.3510101} have identified instances of incorrectly labelled log templates in the dataset. As a result, we adopted the corrected benchmark dataset released by 
Khan et al.~\cite{10.1145/3510003.3510101}, following recent research~\cite{le2023log,khan2023impact}.

\phead{Environment and Implementation.}
Our experiments were conducted on a server with an NVIDIA Tesla V100 GPU using CUDA 11.0. For the fine-tuning process of the model, we used a maximum \texttt{learning rate} of 5e-4 and use the AdamW~\cite{loshchilov2018decoupled} optimizer with a linear learning rate decay schedule for optimization. 
For single system fine-tuning, we uniformly set the \texttt{batch size} to 5 and trained 30 \texttt{epochs} for \logllms. For the cross-system scenario (RQ4), we trained 20 \texttt{epochs} and boost the \texttt{batch size} to 20 in order to shorten the training time. 
We used OpenPrompt~\cite{ding2021openprompt} to fine-tune \tsmall and \tbase. We used LoRA~\cite{hu2021lora} with PEFT v0.3.0 to fine-tune \llama and \chatGLM 
Note that, fine-tuning with 50 data samples took from only a few seconds to less than five minutes for all the studied LLMs, so the cost of few-shot fine-tuning is small.
For log parsing, we set the \texttt{temperature} to 0 and \texttt{num\_beams} to 2 in the generation configuration in order to generate consistent and stable parsed results. Due to the difference in the length of the prompt template, we set the \texttt{max\_length} to 256 (\tsmall and \tbase) and 512 (\llama and \chatGLM) as the generation parameter, respectively.

\phead{Evaluation Metrics for Log Parsing.}
Following prior studies~\cite{10.1145/3510003.3510101,zhu2019tools,le2023log,dai2020logram,liu2022uniparser,nedelkoski2021self}, we use two metrics to evaluate the effectiveness of LLMs in log parsing: Group Accuracy and Parsing Accuracy. 


\noindent{\bf Group Accuracy (GA):} 
Group Accuracy~\cite{zhu2019tools}
does not directly evaluate the correctness of the parsed logs. Instead, GA assesses the accuracy of the automatically grouped logs (e.g., GroupID$_{parsed}$ shown in Figure~\ref{fig1:definition}). 
GA is calculated as the ratio between the number of correctly grouped logs and the total number of logs.
Specifically, GA first groups logs based on the parsed logs (i.e., generated by a log parser), and compares the resulting groups with grouping results from the ground truth (e.g., GroupID$_{gtruth}$).
For instance, 
the log parsing result in Figure~\ref{fig1:definition} has a GA of 4/7. 
Once the raw logs are parsed, they are grouped into different groups, as illustrated in Figure~\ref{fig1:definition}. We can see that Log 1 and 2 are grouped together, Log 6 and 7 are grouped together, and Log 3, 4, and 5 form a separate group by themselves. 
The grouping results (GroupID$_{parsed}$) for group 1 to group 3 match the grouping results obtained by using the ground truth log template (GroupID$_{gtruth}$). 
However, Log 5 to 7 form two groups (GroupID$_{parsed}$ 4 and 5) when using the parsed log template, whereas there is only one group (GroupID$_{gtruth}$ 4) if using the log template from the ground truth. As a result, the GA for this example is 4/7. 

Although GA is commonly used, prior studies~\cite{liu2022uniparser,10.1145/3510003.3510101,dai2020logram} highlighted its limitations. For instance, even if the logs are perfectly grouped with a GA of 100\%, the parsed log template may not fully match the ground truth template due to the misidentified parameters in the parsed templates. As a result, GA cannot show whether or not the logs are parsed correctly. Furthermore, if the logs are not parsed correctly but are still grouped in a cluster (e.g., Log 3 and 4 in Figure~\ref{fig1:definition}), GA will still be 100\%.

\noindent{\bf Parsing Accuracy (PA):} 
Parsing Accuracy~\cite{liu2022uniparser}
within the log template must match with the ground truth template.
For example,
the log parsing result in Figure~\ref{fig1:definition} has a PA of 3/7 because there are missed variables in Log 3, 4, 6, and 7 (highlighted in {\em red}). 
Hence, PA is a stricter metric compared to GA and aligns more closely with practical requirements~\cite{10.1145/3510003.3510101,le2023log,liu2022uniparser}. Prior studies also found that parsing the variables can help downstream log analysis tasks~\cite{khan2023impact,10.1007/978-3-030-88494-9_16,lu2017log}, which further shows the importance of PA over GA. 

\section{Evaluation}
\label{sec:evaluation}


\subsection*{RQ1: What is the accuracy of \logllms? }

\noindent{\bf Motivation.} 
A recent work by Le and Zhang~\cite{le2023log} proposed using Large Language Models (LLMs) to learn from labelled logs and accurately identify parameters in logs. 
However, their study only provided initial evidence on the feasibility of using LLMs for log parsing, as they solely utilized one masked-language LLM (RoBERTa-base). Yet, there is a lack of research exploring the impact of different types (e.g., text2text or text generation LLMs) and sizes of LLMs on log parsing accuracy.
Hence, in this RQ, we aim to investigate the differences among various types of LLMs and the impact of distinct LLM parameter sizes on log parsing. 
Such comparisons can assist practitioners in identifying the most effective LLM for log parsing, while also enabling researchers to identify potential future directions on LLM-based log parsing. 

\begin{table*}
\centering
\caption{A comparison of the grouping accuracy (GA) and parsing accuracy (PA) for the state-of-the-art (first three columns) and the \logllm (the last four columns) parsers. 
}\label{tab:RQ1}
\vspace{-1em}
\scalebox{0.75}{
\setlength{\tabcolsep}{3mm}{
\begin{tabular}{l|cc|cc|cc?cc|cc|cc|cc}\toprule
& \multicolumn{2}{c|}{\textbf{Drain}} & \multicolumn{2}{c|}{\textbf{Logram}} & \multicolumn{2}{c?}{\textbf{LogPPT}} & \multicolumn{2}{c|}{\textbf{\tsmall}} & \multicolumn{2}{c|}{\textbf{\tbase}} & \multicolumn{2}{c|}{\textbf{\llama}} & \multicolumn{2}{c}{\textbf{\chatGLM}}\\

\cline{2-15}
&GA &PA &GA &PA &GA &PA &GA &PA &GA &PA &GA &PA &GA &PA \\
\hline
Android &0.8305 &0.5475 &0.7420 &0.2780 &\textbf{0.8845} &0.7665 &0.8015 &0.9005 &0.8680 &0.9375 &0.8485 &\textbf{0.9455} &0.8315 &0.8395 \\
Apache &\textbf{1} &0.6935 &0.3125 &0.0065 &\textbf{1} &0.9940 &\textbf{1} &\textbf{1} &\textbf{1} &\textbf{1} &\textbf{1} &\textbf{1} &\textbf{1} &\textbf{1} \\
BGL &\textbf{0.9625} &0.3420 &0.5870 &0.1245 &0.9535 &0.9695 &0.5040 &0.9745 &0.4985 &0.9770 &0.9415 &\textbf{0.9805} &0.9440 &0.9640 \\
Hadoop &0.9475 &0.2690 &0.4510 &0.1125 &0.9935 &0.8950 &\textbf{1} &0.9140 &0.8055 &0.9125 &0.9805 &\textbf{0.9825} &0.6440 &0.8375 \\
HDFS &0.9975 &0.3545 &0.9300 &0.0045 &\textbf{1} &0.9025 &\textbf{1} &\textbf{1} &\textbf{1} &\textbf{1} &0.9575 &0.9880 &0.9575 &0.9965 \\
HealthApp &0.7800 &0.2305 &0.2665 &0.1120 &\textbf{1} &0.7885 &0.8025 &0.9560 &0.8085 &0.9010 &0.8550 &\textbf{0.9955} &0.5540 &0.8190 \\
HPC &0.8870 &0.6355 &0.9105 &0.6430 &0.9430 &0.9470 &0.9685 &0.9835 &\textbf{0.9740} &0.9895 &0.9700 &\textbf{0.9935} &0.9700 &0.9855 \\
Linux &0.6900 &0.1835 &0.3610 &0.1240 &\textbf{0.9335} &0.9485 &0.1785 &0.8515 &0.8190 &0.9385 &0.5455 &0.8385 &0.8785 &\textbf{0.9495} \\
Mac &\textbf{0.7865} &0.2175 &0.5680 &0.1685 &0.7800 &0.6725 &0.7325 &0.6725 &0.7750 &\textbf{0.7090} &0.7390 &0.6765 &0.6505 &0.5205 \\
OpenSSH &0.7890 &0.5080 &0.6105 &0.2980 &0.6275 &0.9795 &0.8870 &0.9860 &\textbf{1} &\textbf{1} &0.7095 &0.9935 &0.5840 &0.9450 \\
OpenStack &0.7325 &0.0185 &0.3255 &0 &0.9890 &0.9065 &0.9890 &0.9895 &\textbf{1} &0.9885 &0.9785 &\textbf{0.9960} &0.3125 &0.8725 \\
Proxifier &0.5265 &0 &0.5035 &0 &\textbf{1} &\textbf{1} &\textbf{1} &\textbf{1} &\textbf{1} &\textbf{1} &\textbf{1} &\textbf{1} &0.0310 &0.9150 \\
Spark &0.9200 &0.3595 &0.2820 &0.2585 &\textbf{0.9990} &0.9910 &0.8500 &\textbf{0.9995} &0.8500 &0.9905 &0.9850 &0.9850 &0.7750 &0.9585 \\
Thunderbird &0.9550 &0.0465 &0.5540 &0.0040 &0.6790 &0.9255 &0.9630 &0.9595 &\textbf{0.9705} &\textbf{0.9730} &0.6925 &0.9675 &0.9560 &0.9375 \\
Windows &0.9970 &0.4620 &0.6940 &0.1405 &0.9910 &0.9830 &0.7155 &0.9885 &0.7155 &0.9950 &\textbf{0.9985} &\textbf{0.9965} &0.9920 &0.9880 \\
Zookeeper &0.9665 &0.4970 &0.7235 &0.4735 &0.9935 &0.9895 &0.9945 &\textbf{0.9995} &\textbf{1} &\textbf{0.9995} &0.9945 &\textbf{0.9995} &0.9935 &0.9930 \\
\hline
Average &0.8605 &0.3353 &0.5513 &0.1718 &\textbf{0.9229} &0.9162 &0.8367 &0.9484 &0.8803 &0.9570 &0.8873 &\textbf{0.9587} &0.7546 &0.9076 \\
\bottomrule
\end{tabular}}}
   \begin{tablenotes}
     \item \footnotesize{Note: The highest values of GA and PA for each system are highlighted in \textbf{bold}. The results of Drain and Logram are based on the evaluation conducted by Khan et al.~\cite{10.1145/3510003.3510101} on the corrected log dataset.} 
   \end{tablenotes}
\vspace{-1em}
\end{table*}

\noindent{\bf Approach.} 
For each system, we fine-tune the four \logllms 
using 50 logs sampled using the sampling algorithm (Algorithm~\ref{alg:sample}). 
Then, similar to prior studies~\cite{le2023log,li2023did}, we use each fine-tuned model to generate the log templates for all 2,000 logs for each of the 16 systems.
We compare the grouping and parsing accuracy against state-of-the-art approaches: Drain~\cite{Drain}, Logram~\cite{dai2020logram}, and LogPPT~\cite{le2023log}. 
We selected the state-of-the-art approaches based on their high accuracy and efficiency~\cite{zhu2019tools, dai2020logram}. 

\phead{Results.} 
\noindent{\bf {\em \logllms have a higher (4.25\% to 78.69\% higher for \llama) parsing accuracy compared to state-of-the-arts log parsers.}}
Table~\ref{tab:RQ1} shows both the grouping accuracy (GA) and parsing accuracy (PA) of the state-of-the-art and \logllms. 
We find that, in general, \logllms have a higher GA (0.7546\textasciitilde0.8873) and PA (0.9076\textasciitilde0.9587) compared to the traditional log parsers (GA of 0.5513 and 0.8605, and PA of 0.3353 and 0.1718, for Drain and Logram, respectively). 
LLMs such as \llama and \tbase achieve a GA of 0.88 and a PA of almost 0.96.  
LogPPT, on the other hand, achieves a high GA (0.9229) and PA (0.9162). 
Compared to LogPPT which uses a masked language model, we find that our parsers that are based on text2text and text generation models achieve a better PA (except for \chatGLM). For example, \llama achieves a PA of 0.9587, which is 4.6\% higher than that of LogPPT. 



\noindent{\bf {\em The probabilistic nature of text generation and text2text LLMs may have an impact on the grouping accuracy (GA) of \logllms. Nevertheless, the differences in GA between LogPPT and two \logllms are not statistically significant. }} 
Unlike traditional algorithms, 
text generation and text2text models have a small probability of generating erratic output that leads to parsing errors~\cite{BELFORD2018159}, which has a larger impact on GA. 
For example, we observed that when we parsed 2,000 logs from Spark using \tsmall, 1,999 logs were parsed correctly, resulting in a PA of 0.9995.
Only one log was parsed incorrectly (one of the dynamic variables was not parsed correctly).
However, this log shares the same template with 299 other logs in the system. Although 299/300 logs were correctly parsed, these 300 logs were not grouped together because that one incorrectly parsed log forms a group by itself. 
As a result, the GA for Spark becomes 0.85 (1,700/2,000). 
Note that, we set the temperature of the \logllms to zero to ensure the consistency in the parsed logs (i.e., given the same input prompt, the output will be the same)~\cite{wang2023cost}. However, logs contain dynamically generated values, so even if two logs have the same template, they are considered two distinct inputs and may have a small probability of resulting in parsed logs with small differences. 

{\em Nevertheless, compared to state-of-the-art parsers, the PA of both \tbase and \llama showed a statistically significant increase using paired t-test (p-value<0.05), while GA did not exhibit a statistically significant difference compared to LogPPT (p-value>0.05).} 
Prior studies~\cite{10.1145/3510003.3510101, dai2020logram,le2023log} stated that PA evaluates the practical goal (i.e., correctly parsing logs) of log parsing, which makes PA a better evaluation metric than GA.
In short, \logllms can better identify the variables of logs and generate correct log templates matching the ground truth.

\begin{table}\centering
\setlength{\extrarowheight}{1pt}
\caption{Grouping accuracy (GA) and parsing accuracy (PA) for logs outside the training dataset.}\label{tab: remove}
\vspace{-1em}
\scalebox{0.7}{
\begin{tabular}{p{1.5cm}|cc|cc|cc|cc}\toprule
&\multicolumn{2}{p{2.4cm}<{\centering}|}{\textbf{\tsmall}} & \multicolumn{2}{p{2.2cm}<{\centering}|}{\textbf{\tbase}} & \multicolumn{2}{p{2.3cm}<{\centering}|}{\textbf{\llama}} & \multicolumn{2}{p{2.3cm}<{\centering}}{\textbf{\chatGLM}} \\\cline{2-9}
&\textbf{GA} &\textbf{PA} &\textbf{GA} &\textbf{PA} &\textbf{GA} &\textbf{PA} &\textbf{GA} &\textbf{PA} \\\hline
Android &0.7716 &0.8452 &0.9058 &0.8647 &0.8789 &0.8680 &0.8148 &0.8329 \\
Apache &1 &1 &1 &1 &1 &1 &1 &1 \\
BGL &0.3798 &0.9720 &0.3709 &0.9752 &0.9349 &0.9758 &0.9375 &0.9585 \\
Hadoop &1 &0.8517 &0.6914 &0.8526 &0.9768 &0.9662 &0.3443 &0.8501 \\
HDFS &1 &1 &1 &1 &0.9605 &0.9892 &0.9605 &0.9979 \\
HealthApp &0.7547 &0.9457 &0.7622 &0.8770 &0.8192 &0.9944 &0.4426 &0.7765 \\
HPC &0.9040 &0.9584 &0.9200 &0.9744 &0.9154 &0.9789 &0.9106 &0.9610 \\
Linux &0.0474 &0.8456 &0.9465 &0.9569 &0.4392 &0.7996 &0.8834 &0.9504 \\
Mac &0.6583 &0.5782 &0.7106 &0.6247 &0.6651 &0.5857 &0.5678 &0.4335 \\
OpenSSH &0.8500 &0.9848 &1 &1 &0.5129 &0.9889 &0.3036 &0.9082 \\
OpenStack &0.9865 &0.9878 &1 &0.9865 &0.9737 &0.9955 &0.1432 &0.8504 \\
Proxifier &1 &1 &1 &1 &1 &1 &0.0137 &0.8901 \\
Spark &0.8443 &0.9995 &0.8443 &0.9901 &0.9854 &0.9854 &0.7695 &0.9629 \\
Thunderbird &0.9550 &0.9498 &0.9648 &0.9667 &0.6638 &0.9579 &0.9487 &0.9303 \\
Windows &0.5271 &0.9825 &0.5271 &0.9925 &0.9983 &0.9975 &0.9882 &0.9891 \\
Zookeeper &0.9886 &0.9989 &1 &1 &0.9886 &0.9989 &0.9886 &0.9874 \\
\hline
Average &0.7917 &0.9313 &0.8527 &0.9413 &0.8570 &0.9426 &0.6886 &0.8925 \\
\bottomrule
\end{tabular}}
\end{table}

\noindent{\bf {\em After we remove the logs used for few-shot fine-tuning during evaluation, the average GA and PA of \logllm on the remaining logs decrease slightly but are still higher than other baselines.}} Previous studies on log parsers usually evaluate and compare the effectiveness of log parsers on the entire data set (i.e., all 2,000 logs)~\cite{Drain,le2023log,zhu2019tools}. However, such evaluation approaches may include the training data in the evaluation process, causing potential data leakage issues. Hence, we re-evaluated the GA and PA of all four \logllms only on the logs 
by \textbf{removing all the logs from the test set that were exactly the same as those in the training set} and showing the results in Table~\ref{tab: remove}. 
Compared with the baseline result in Table~\ref{tab:RQ1} (evaluated using all the logs), all four \logllms’ average GA and average PA experienced a slight decrease, with the average PA decreasing by less than 1\%. Nevertheless, \tbase and \llama still have higher average PAs (0.9413 for \tbase and 0.9426 for \llama) compared to LogPPT (PA is 0.9162, but LogPPT included the training data in the evaluation, having a potential data leakage issue). Our findings show that, after removing the log samples used for training in the evaluation process, \logllms can still achieve higher PA than all the baselines. 

\noindent{\bf {\em While more complex \logllms (more parameters) generally give better parsing results, simpler models already give promising results. Future studies should consider the trade-off between parsing accuracy and the complexity of LLMs. }} In general, more complex models give better parsing results, with the exception of \chatGLM. 
One reason that \chatGLM has a worse accuracy may be that it is a bilingual model and the logs are written in English. Another possible reason is that ChatGLM is engineered as a chatbot, which is fine-tuned to give human-like responses. 
Between \tsmall and \tbase, the GA and PA increased by 5.21\% and 0.9\%, respectively. We see a further improvement when using \llama compared to \tbase, although the improvement is small (0.8\% and 0.17\% in GA and PA). 

However, more complex models may require more resources and time to parse the logs. We randomly selected 100 logs from Spark's log dataset, and measured the average parsing time by repeating the process 20 times. 
Under the same hardware environment, complex \logllms require a longer parsing time. \tsmall and \tbase could parse 100 logs in an average of 1.27 seconds and 4 seconds, respectively, while \chatGLM and \llama require 19.87 and 28.93 seconds, respectively. 
Therefore, future studies should encompass a trade-off between the accuracy and efficiency of using LLMs on log parsing.

\rqboxc{\logllms achieve better PA than state-of-the-art and similar GA compared to LogPPT. Although \llama, which has a larger number of model parameters, achieves the best GA and PA, the difference is small compared to smaller LLMs like \tbase.}

\begin{table*}[!htp]\centering

\setlength{\extrarowheight}{1pt}
\caption{Grouping accuracy (GA) and parsing accuracy (PA) for different numbers of training shots.
}\label{tab: RQ2}
\vspace{-1em}

\scalebox{0.75}{
{\setlength{\tabcolsep}{0.5mm}{
\begin{tabular}{l|cc|cc|cc|cc?cc|cc|cc|cc?cc|cc|cc|cc?cc|cc|cc|cc}\toprule
&\multicolumn{8}{c?}{\textbf{\tsmall}} &\multicolumn{8}{c?}{\textbf{\tbase}} &\multicolumn{8}{c?}{\textbf{\llama}}&\multicolumn{8}{c}{\textbf{\chatGLM}} \\
\cline{2-33}
&\multicolumn{2}{c|}{25 shots} &\multicolumn{2}{c|}{50 shots} &\multicolumn{2}{c|}{75 shots} &\multicolumn{2}{c?}{100 shots} &\multicolumn{2}{c|}{25 shots} &\multicolumn{2}{c|}{50 shots} &\multicolumn{2}{c|}{75 shots} &\multicolumn{2}{c?}{100 shots} &\multicolumn{2}{c|}{25 shots} &\multicolumn{2}{c|}{50 shots} &\multicolumn{2}{c|}{75 shots} &\multicolumn{2}{c?}{100 shots} &\multicolumn{2}{c|}{25 shots} &\multicolumn{2}{c|}{50 shots} &\multicolumn{2}{c|}{75 shots} &\multicolumn{2}{c}{100 shots} \\\cline{2-33}
&GA &PA &GA &PA &GA &PA &GA &PA &GA &PA &GA &PA &GA &PA &GA &PA &GA &PA &GA &PA &GA &PA &GA &PA &GA &PA &GA &PA &GA &PA &GA &PA \\\hline
Android & 0.90& 0.88& 0.80& 0.90& \textbf{0.97}& \textbf{0.93}& 0.86& \textbf{0.93}& 0.95& 0.91& 0.87& 0.94& \textbf{0.98}& 0.96& 0.96& \textbf{0.97}& 0.95& 0.91& 0.85& 0.95& \textbf{0.98}& 0.98& 0.97& \textbf{0.99}& 0.78& 0.68& 0.83& 0.84& \textbf{0.93}& \textbf{0.96}& 0.83& 0.93\\
Apache & \textbf{1}& \textbf{1}& \textbf{1}& \textbf{1}& \textbf{1}& \textbf{1}& \textbf{1}& \textbf{1}& \textbf{1}& \textbf{1}& \textbf{1}& \textbf{1}& \textbf{1}& \textbf{1}& \textbf{1}& \textbf{1}& \textbf{1}& \textbf{1}& \textbf{1}& \textbf{1}& \textbf{1}& \textbf{1}& \textbf{1}& \textbf{1}& \textbf{1}& 0.99& \textbf{1}& \textbf{1}& \textbf{1}& \textbf{1}& \textbf{1}& \textbf{1}\\
BGL & \textbf{0.56}& 0.93& 0.50& 0.97& 0.51& \textbf{0.98}& \textbf{0.56}& \textbf{0.98}& \textbf{0.60}& 0.94& 0.50& 0.98& \textbf{0.60}& 0.98& \textbf{0.60}& \textbf{0.99}& 0.74& 0.84& 0.94& 0.98& 0.85& \textbf{0.99}& \textbf{0.96}& \textbf{0.99}& 0.84& 0.90& 0.94& 0.96& 0.35& 0.92& \textbf{0.96}& \textbf{0.98}\\
Hadoop & 0.80& 0.87& \textbf{1}& 0.91& 0.98& 0.90& 0.97& \textbf{0.92}& 0.99& 0.89& 0.81& 0.91& \textbf{1}& \textbf{0.99}& 0.97& \textbf{0.99}& \textbf{0.99}& 0.96& 0.98& 0.98& 0.98& \textbf{0.99}& 0.96& 0.98& 0.95& 0.61& 0.64& 0.84& \textbf{0.96}& \textbf{0.97}& 0.79& 0.83\\
HDFS & 0.70& 0.98& \textbf{1}& \textbf{1}& \textbf{1}& \textbf{1}& \textbf{1}& \textbf{1}& \textbf{1}& \textbf{1}& \textbf{1}& \textbf{1}& \textbf{1}& \textbf{1}& \textbf{1}& \textbf{1}& 0.92& 0.92& 0.96& 0.99& 0.96& \textbf{1}& \textbf{1}& \textbf{1}& 0.84& 0.94& \textbf{0.96}& \textbf{1}& 0.84& 0.99& 0.80& 0.99\\
HealthApp & 0.67& 0.77& \textbf{0.80}& \textbf{0.96}& \textbf{0.80}& 0.93& 0.67& 0.94& 0.66& 0.85& \textbf{0.81}& 0.90& \textbf{0.81}& \textbf{0.99}& \textbf{0.81}& \textbf{0.99}& 0.86& 0.99& 0.86& \textbf{1}& 0.87& \textbf{1}& \textbf{1}& \textbf{1}& 0.66& 0.73& 0.55& 0.82& 0.68& \textbf{0.86}& \textbf{0.73}& \textbf{0.86}\\
HPC & 0.95& 0.98& 0.97& 0.98& 0.97& \textbf{0.99}& \textbf{0.98}& \textbf{0.99}& 0.97& 0.98& 0.97& 0.99& 0.97& 0.99& \textbf{1}& \textbf{1}& \textbf{0.99}& 0.99& 0.97& 0.99& 0.98& \textbf{1}& \textbf{0.99}& \textbf{1}& 0.77& 0.91& \textbf{0.97}& \textbf{0.99}& 0.95& \textbf{0.99}& 0.93& \textbf{0.99}\\
Linux & \textbf{0.36}& \textbf{0.89}& 0.18& 0.85& \textbf{0.36}& \textbf{0.89}& 0.18& 0.87& \textbf{0.94}& \textbf{0.97}& 0.82& 0.94& 0.88& 0.91& 0.36& 0.89& \textbf{1}& \textbf{0.99}& 0.55& 0.84& 0.76& 0.98& 0.24& 0.95& 0.42& 0.87& \textbf{0.88}& 0.95& 0.48& \textbf{0.96}& 0.69& 0.92\\
Mac & 0.64& 0.52& 0.73& 0.67& 0.77& 0.76& \textbf{0.82}& \textbf{0.80}& 0.70& 0.59& 0.78& 0.71& 0.82& 0.73& \textbf{0.84}& \textbf{0.83}& 0.77& 0.58& 0.74& 0.68& 0.79& 0.78& \textbf{0.82}& \textbf{0.80}& 0.70& 0.44& 0.65& 0.52& 0.70& 0.53& \textbf{0.83}& \textbf{0.79}\\
OpenSSH & 0.63& 0.92& 0.89& 0.99& 0.64& 0.98& \textbf{0.94}& \textbf{1}& 0.33& 0.88& \textbf{1}& \textbf{1}& 0.81& \textbf{1}& \textbf{1}& \textbf{1}& 0.44& 0.99& 0.71& 0.99& \textbf{0.94}& \textbf{1}& 0.75& \textbf{1}& 0.58& 0.90& 0.58& 0.95& 0.58& 0.98& \textbf{0.75}& \textbf{0.99}\\
OpenStack & 0.96& 0.94& \textbf{0.99}& 0.99& \textbf{0.99}& \textbf{1}& \textbf{0.99}& 0.99& 0.97& 0.95& \textbf{1}& 0.99& 0.99& \textbf{1}& \textbf{1}& \textbf{1}& 0.52& 0.84& 0.98& \textbf{1}& 0.99& \textbf{1}& \textbf{1}& \textbf{1}& 0.31& 0.51& 0.31& 0.87& 0.47& 0.83& \textbf{0.99}& \textbf{1}\\
Proxifier & 0.53& \textbf{1}& \textbf{1}& \textbf{1}& \textbf{1}& \textbf{1}& \textbf{1}& \textbf{1}& \textbf{1}& \textbf{1}& \textbf{1}& \textbf{1}& \textbf{1}& \textbf{1}& \textbf{1}& \textbf{1}& 0.52& \textbf{1}& \textbf{1}& \textbf{1}& 0.05& \textbf{1}& 0.53& \textbf{1}& 0.50& 0.95& 0.03& 0.92& 0.53& \textbf{1}& \textbf{0.98}& \textbf{1}\\
Spark & 0.36& 0.95& \textbf{0.85}& \textbf{1}& \textbf{0.85}& \textbf{1}& 0.66& \textbf{1}& 0.85& \textbf{1}& 0.85& 0.99& \textbf{1}& \textbf{1}& \textbf{1}& \textbf{1}& 0.81& \textbf{1}& 0.99& 0.99& \textbf{1}& 0.99& \textbf{1}& \textbf{1}& 0.78& 0.79& 0.78& 0.96& 0.79& \textbf{0.97}& \textbf{0.87}& 0.94\\
Thunderbird & \textbf{0.96}& 0.93& \textbf{0.96}& 0.96& 0.68& 0.97& 0.68& \textbf{0.98}& 0.96& 0.92& \textbf{0.97}& 0.97& 0.70& \textbf{0.98}& 0.70& \textbf{0.98}& 0.68& 0.96& \textbf{0.69}& \textbf{0.97}& 0.68& \textbf{0.97}& 0.68& \textbf{0.97}& 0.64& 0.57& \textbf{0.96}& 0.94& 0.67& \textbf{0.95}& 0.67& \textbf{0.95}\\
Windows & 0.71& 0.98& 0.72& 0.99& \textbf{1}& \textbf{1}& \textbf{1}& \textbf{1}& 0.99& 0.99& 0.72& \textbf{1}& \textbf{1}& \textbf{1}& \textbf{1}& \textbf{1}& \textbf{1}& \textbf{1}& \textbf{1}& \textbf{1}& \textbf{1}& \textbf{1}& \textbf{1}& \textbf{1}& \textbf{0.99}& 0.84& \textbf{0.99}& \textbf{0.99}& \textbf{0.99}& \textbf{0.99}& \textbf{0.99}& \textbf{0.99}\\
Zookeeper & \textbf{1}& 0.99& 0.99& \textbf{1}& 0.99& \textbf{1}& \textbf{1}& \textbf{1}& 0.99& \textbf{1}& \textbf{1}& \textbf{1}& 0.99& \textbf{1}& 0.99& \textbf{1}& \textbf{0.99}& 0.99& \textbf{0.99}& \textbf{1}& \textbf{0.99}& \textbf{1}& \textbf{0.99}& \textbf{1}& \textbf{0.99}& \textbf{0.99}& \textbf{0.99}& \textbf{0.99}& \textbf{0.99}& \textbf{0.99}& 0.97& 0.97\\
\hline
Average & 0.73& 0.91& \textbf{0.84}& 0.95& \textbf{0.84}& \textbf{0.96}& 0.83& \textbf{0.96}& 0.87& 0.93& 0.88& 0.96& \textbf{0.91}& 0.97& 0.89& \textbf{0.98}& 0.82& 0.93& \textbf{0.89}& 0.96& 0.86& \textbf{0.98}& 0.87& \textbf{0.98}& 0.73& 0.79& 0.75& 0.91& 0.75& 0.93& \textbf{0.86}& \textbf{0.95}\\

\bottomrule
\end{tabular}}}}
   \begin{tablenotes}
     \item \footnotesize{Note: The highest values of GA and PA for each LLM for each system are highlighted in \textbf{bold}.}
   \end{tablenotes}
\vspace{-1.5em}
\end{table*}

\subsection*{RQ2: How does the accuracy of log parsing vary under different shot sizes?}

\noindent{\bf Motivation.} 
In RQ1, we ascertain that LLMs exhibit superior accuracy in log parsing compared to the state-of-the-art approaches. 
When using LLMs, one thing that researchers and practitioners need to decide is the number of samples for few-shot tuning. 
Prior research~\cite{gao-etal-2021-making,BAILLY2022106504} has demonstrated that the efficacy of a model fine-tuned for an individual task is contingent upon the size and diversity of the training dataset.
However, manually labelling data can be time-consuming and manual-intensive. Hence, in this RQ, we examine the ramifications of varying training set sizes on the accuracy of log parsing tasks when employing distinct LLMs. 

\noindent{\bf Approach.} 
For each system, we sample 25, 50, 75, and 100 log lines and their corresponding log template using our log sampling algorithm (Algorithm~\ref{alg:sample}). The same sets of logs are used as the fine-tuning dataset for all \logllms.
We then evaluate the log parsing performance (i.e., GA and PA) of \logllms using different sizes of fine-tuning datasets.
We vertically compare the accuracy changes of each \logllm after increasing the training data size. Simultaneously, we also horizontally compare the accuracy differences of different \logllms under the same training data size.

To compare, we also apply in-context learning on \tbase and \llama without fine-tuning to investigate the log parsing accuracy. We chose these two LLMs because of their large size and good parsing results shown in RQ1. 
We use 3 and 15 shots (in-context log parsing demonstrations), respectively for the two LLMs, due to their limits on the size of the input tokens.

\phead{Results.} 
\noindent{\bf {\em Increasing the number of shots increases the accuracy of \logllms, but the difference is small (e.g., 1--2\%) or fluctuates for most \logllms beyond 50 shots, except for \chatGLM.}}
Table~\ref{tab: RQ2} shows the accuracy of \logllms using different numbers of shots.
Although there are some improvements in PA and GA when the number of shots increases, the accuracies stabilize after 50 shots. 
When the shot size is 25, both GA and PA decrease (1\% to 15\% and 3\% to 15\%, respectively) for all \logllms compared to using 50 shots. 
We also find that \logllms have different sensitivity on the shot sizes. For example, \tbase is relatively stable across all shot sizes, while \chatGLM has the largest improvement when the shot size increases. 
When the shot size is increased to 100, the GA decreases for all \logllms except for \chatGLM, but the improvement for PA is barely noticeable. We also find that \tbase has better GA and the same PA compared to \llama, the largest LLM among the four studied LLMs. Our finding shows that more complex LLMs may not achieve better PA and GA.
For instance, the smallest model \tsmall achieves comparable results to the second largest model \chatGLM, and their difference remains trivial even with the increased training shots.

\begin{table*}\centering
\setlength{\extrarowheight}{1pt}

\caption{Grouping accuracy (GA) and parsing accuracy (PA) for different shots of in-context learning.
}\label{tab: icl}
\vspace{-1em}
\scalebox{0.85}{
\begin{tabular}{l|cc|cc|cc|cc?cc|cc|cc|cc|cc}\toprule
&\multicolumn{8}{c?}{\textbf{\llama}} &\multicolumn{10}{c}{\textbf{\tbase}}\\
\cline{2-19}
&\multicolumn{2}{c|}{\textbf{5 shots}} &\multicolumn{2}{c|}{\textbf{10 shots}} &\multicolumn{2}{c|}{\textbf{15 shots}} &\multicolumn{2}{c?}{\textbf{20 shots}} &\multicolumn{2}{c|}{\textbf{1 shots}} &\multicolumn{2}{c|}{\textbf{2 shots}} &\multicolumn{2}{c|}{\textbf{3 shots}} &\multicolumn{2}{c|}{\textbf{4 shots}} &\multicolumn{2}{c}{\textbf{5 shots}} \\\cline{2-19}
&GA &PA &GA &PA &GA &PA &GA &PA &GA &PA &GA &PA &GA &PA &GA &PA &GA &PA \\\hline
Android &0.200 &0.093 &0.393 &0.126 &0.430 &0.222 &0.642 &0.318 &0.423 &0.043 &0.427 &0.018 &0.424 &0.002 &0.482 &0.033 &0.530 &0.021 \\
Apache &0.984 &0.700 &0.430 &0.460 &0.709 &0.711 &1 &0.725 &0.291 &0 &0.291 &0 &0.566 &0 &0.291 &0 &0.291 &0 \\
BGL &0.240 &0.118 &0.258 &0.211 &0.368 &0.418 &0.238 &0.665 &0.126 &0 &0.085 &0.001 &0.130 &0.002 &0.133 &0.009 &0.154 &0.008 \\
Hadoop &0.335 &0.244 &0.407 &0.395 &0.180 &0.164 &0.255 &0.321 &0.116 &0 &0.188 &0.003 &0.213 &0.003 &0.456 &0.004 &0.284 &0.006 \\
HDFS &0.001 &0.071 &0.041 &0.178 &0.011 &0.335 &0.001 &0.241 &0.001 &0 &0.001 &0 &0.001 &0 &0.001 &0 &0.001 &0 \\
HealthApp &0.332 &0.561 &0.429 &0.528 &0.626 &0.756 &0.584 &0.824 &0.120 &0.001 &0.127 &0 &0.128 &0 &0.147 &0.019 &0.074 &0.012 \\
HPC &0.295 &0.490 &0.155 &0.506 &0.354 &0.592 &0.244 &0.478 &0.635 &0.001 &0.384 &0.001 &0.595 &0 &0.390 &0 &0.530 &0.081 \\
Linux &0.036 &0.106 &0.201 &0.611 &0.264 &0.616 &0.199 &0.675 &0.129 &0.002 &0.167 &0.009 &0.154 &0.012 &0.170 &0.024 &0.165 &0.012 \\
Mac &0.247 &0.129 &0.345 &0.155 &0.414 &0.201 &0.384 &0.180 &0.247 &0.012 &0.355 &0.066 &0.344 &0.008 &0.378 &0.024 &0.389 &0.056 \\
OpenSSH &0.002 &0.182 &0.068 &0.382 &0.074 &0.361 &0.028 &0.389 &0.095 &0 &0.025 &0 &0.255 &0 &0.076 &0 &0.094 &0 \\
OpenStack &0.041 &0 &0.049 &0.054 &0.083 &0.025 &0.070 &0.012 &0.194 &0 &0.147 &0 &0.102 &0 &0.112 &0 &0.097 &0 \\
Proxifier &0.050 &0.627 &0.050 &0.981 &0.050 &0.980 &0.050 &0.963 &0 &0 &0.001 &0 &0 &0 &0 &0 &0.003 &0 \\
Spark &0.003 &0.312 &0.023 &0.440 &0.401 &0.579 &0.215 &0.494 &0.009 &0 &0.023 &0.002 &0.048 &0.003 &0.064 &0.001 &0.027 &0.003 \\
Thunderbird &0.079 &0.062 &0.118 &0.326 &0.165 &0.398 &0.019 &0 &0.203 &0.010 &0.178 &0.004 &0.168 &0.004 &0.123 &0.004 &0.101 &0.004 \\
Windows &0.162 &0.003 &0.398 &0.567 &0.410 &0.427 &0.181 &0.309 &0.033 &0 &0.132 &0 &0.259 &0.009 &0.139 &0.006 &0.011 &0 \\
Zookeeper &0.171 &0.142 &0.017 &0.206 &0.294 &0.380 &0.248 &0.469 &0.171 &0.158 &0.152 &0.001 &0.173 &0.133 &0.042 &0.020 &0.037 &0.027 \\
\hline
Average &0.198 &0.240 &0.211 &0.383 &0.302 &0.448 &0.272 &0.441 &0.174 &0.014 &0.167 &0.006 &0.222 &0.011 &0.187 &0.009 &0.174 &0.014 \\
\bottomrule
\end{tabular}}
\vspace{-1.3em}
\end{table*}

\noindent{\bf {\em Compared
to in-context learning, few-shot tuning achieves better parsing
accuracy}}. 
Table~\ref{tab: icl} shows the accuracy for different shots of in-context learning.
However, in-context learning yielded worse results. \llama achieved only an average GA and PA of 30\% and 45\%, respectively, across 16 systems. \tbase also performed poorly, resulting in an average GA and PA of only 22\% and 1\%. The finding aligns with the discussion provided by a recent study~\cite{mosbach2023few} and Flan-T5 developers~\cite{flan-t5} where few-shot tuning achieves better results than in-context learning.


\noindent{\bf {\em Different \logllms may require different shot sizes to achieve good accuracy in log parsing. }}
For example, \tbase already achieves very high accuracy (87\% and 93\% for GA and PA) when fine-tuned using only 25 shots. However, as the number of shots increases further, the improvement consistently plateaus among all systems, especially when the shot size is over 75.
On the other hand, when the shot size is 100, the accuracy of \chatGLM becomes comparable to the other parsers.
This trend is particularly noticeable in OpenStack and Proxifier, where both GA and PA show substantial improvements from 25 shots to 100 shots (GA increases from 0.31 to 0.99, and PA increases from 0.51 to 1 for OpenStack, and GA increases from 0.50 to 0.98, and PA increases from 0.95 to 1 for Proxifier).

The variation in the accuracy of log parsing across different \logllms may be attributed to the nature of LLMs as statistical models. Each model learns to parse logs by identifying distinct patterns from the training shots. 
To determine the best shot size and reduce manual effort on data creation, future studies should investigate the relationship between the characteristics of the LLMs (e.g., architecture and training data) and the needed data to fine-tune the LLMs for log parsing. 

\rqboxc{For all \logllms except for \chatGLM, the accuracy improvement in log parsing becomes small or starts to fluctuate when the shot size is over 50. Different LLMs may also require different shot sizes to achieve good parsing results, and more shots do not always give the best results. }

\vspace{-0.5em} 
\subsection*{RQ3:  How is the generalizability of \logllms on unseen log templates?}


\noindent{\bf Motivation.} 
In RQ2, we studied the accuracy of \logllms using various numbers of shots. We found that although GA and PA improve noticeably when the shot size increases from 25 to 50, the improvement is small or remains almost the same when the shot size is 50 or larger. One hypothesis is that the effectiveness of few-shot tuning is constrained by the diversity of the sampling algorithm when presented with diverse log templates. 
Logs with different variable values may still have the same log template. By including a single log in the fine-tuning process, it is possible to enhance the parsing accuracy for all other logs that share the same log template.  
In other words, 50 shots may not capture all the unique log templates, leading to a saturation point where the LLM fails to generalize well to unseen examples. 
Therefore, our objective is to investigate the extent to which few-shot tuning can generalize to unseen log templates. The finding of this RQ may help future research further improve the accuracy of LLM-based log parsers.

\noindent{\bf Approach.} We use the same set of \logllms that are trained using 50 log samples from prior RQs. 
Specifically, we want to study if a \textbf{\em log's log template was not included in the training}, would such a log have lower parsing accuracy. 
We first identify the log templates and the corresponding logs that were not used for few-shot tuning (we call them log$_{unseen}$). Then, we calculate the PA for log$_{unseen}$ and compare it with the PA for log$_{seen}$. 

\begin{table*}\centering
\caption{\logllms' PA on log$_{unseen}$ and PA on log$_{seen}$ when using 50 log samples. Note that we excluded the systems where all the unique log templates were included in the shots. 
}\label{tab:RQ3}
\vspace{-1em}

\scalebox{0.75}{
\setlength{\tabcolsep}{2.3mm}{
\begin{tabular}{l|r|r|r?cc|cc|cc|cc}\toprule
& & & &\multicolumn{2}{c|}{\textbf{\tsmall}} & \multicolumn{2}{c|}{\textbf{\tbase}} & \multicolumn{2}{c|}{\textbf{\llama}} & \multicolumn{2}{c}{\textbf{\chatGLM}} \\
\cline{5-12}
&Total Template &Unseen Template &Unseen Log &PA-unseen &PA-seen &PA-unseen &PA-seen &PA-unseen &PA-seen &PA-unseen &PA-seen \\
\hline
Android &158 &108 &224 &0.5536 &0.9443 &0.5893 &0.9814 &0.7009 &0.9764 &0.5446 &0.8767 \\
BGL &120 &70 &105 &0.7714 &0.9858 &0.8476 &0.9842 &0.7714 &0.9921 &0.6762 &0.9799 \\
Hadoop &114 &64 &64 &0.5469 &0.9261 &0.6719 &0.9205 &0.5156 &0.9979 &0.3906 &0.8523 \\
HealthApp &75 &25 &25 &0.7200 &0.9590 &0.7200 &0.9033 &0.8000 &0.9980 &0.8400 &0.8187 \\
Linux &116 &66 &66 &0.5152 &0.8630 &0.5909 &0.9504 &0.8182 &0.8392 &0.5909 &0.9617 \\
Mac &341 &291 &819 &0.3297 &0.9102 &0.3993 &0.9238 &0.3761 &0.8848 &0.3907 &0.6105 \\
Thunderbird &149 &99 &103 &0.5146 &0.9837 &0.6505 &0.9905 &0.5728 &0.9889 &0.5631 &0.9578 \\
\hline
Average &153 (Sum: 1073)&103 (Sum: 723)&201 (Sum: 1406)&0.5645 &0.9389 &0.6385 &0.9506 &0.6507 &0.9539 &0.5709 &0.8654 \\
\bottomrule
\end{tabular}}}
\vspace{-0.3cm}
\end{table*}

\phead{Results.} \noindent{\bf {\em The PA on log$_{unseen}$ are much lower (e.g., 0.638 for \llama) compared to the PA on log$_{seen}$ (e.g., 0.9539 for \llama). Although only 4.4\% of the logs have unseen log templates, they account for 50\% of the incorrectly parsed logs.}} 
Table~\ref{tab:RQ3} shows the number of total templates, the number of unseen log templates, the number of log$_{unseen}$, and the PA of log$_{unseen}$ and PA of log$_{seen}$. 
We find that the PA decreases significantly for the log$_{unseen}$ compared to the PA for log$_{seen}$. For example, as shown in Table~\ref{tab:RQ3}, the PAs for log$_{seen}$ of \tbase and \llama are 0.9506 and 0.9539, whereas their average PAs for the log$_{unseen}$ are 0.6385 and 0.6507, respectively. 
After some investigation, we find that around 50\% of the incorrectly parsed logs among all the 16 systems belong to one of the unseen log templates, and the finding is consistent across all LLMs. 
Given that, only 4.4\% of the logs are log$_{unseen}$ (1,406 out of all 32,000 logs from all the systems), they are disproportionately more likely to be parsed incorrectly. 
Hence, our finding suggests that log$_{unseen}$ is one of the bottlenecks to further improving parsing accuracy and shed light on future research in log parsers.  
Nevertheless, we find that \logllms still achieve better PA when parsing log$_{unseen}$ compared to traditional state-of-the-art approaches such as Drain and Logram, which have an average PA of 0.3353 and 0.1718, respectively. 

We observe a decrease in the performance of \logllms when they encounter unseen log templates, indicating their limited ability to generalize.
This behaviour in \logllms may be attributed to the limitation of the training data during fine-tuning, which primarily focuses on identifying seen log templates. 
This limitation becomes more apparent when log templates share high similarities. For example, two logs with similar log templates, such as ``\texttt{(<*>) CMD (<*> <*>)}'' and ``\texttt{(<*>) CMD (run-parts <*>)''}, might be mistakenly parsed as the same log template due to their textual similarity, resulting in reduced accuracy. However, if both logs and their templates were provided as training data, \logllms could better differentiate between them and achieve higher accuracy.
Future research may consider improving the generalization of \logllms by proposing sampling algorithms that can select a more diverse sampled set of logs and templates during fine-tuning. 

\rqboxc{\logllms achieves bad results on log$_{unseen}$ compared with results on log$_{seen}$. The unseen logs, which only make up 4.4\% of all logs, form 50\% of the incorrectly parsed logs.
Some types of variables may not be identified even if they appear in the training dataset. }

\vspace{-1em}
\subsection*{RQ4: Can pre-trained \logllms help improve parsing accuracy?}

\noindent{\bf Motivation.} 
In the previous RQs, we investigated the log parsing accuracy when fine-tuning the LLMs using the logs from the same system. However, one major advantage of LLM is its ability to generalize on new datasets~\cite{9609166,wang2021language,8952475}. Therefore, in this RQ, we study if using a \logllm that is pre-trained using logs from other systems can further improve log parsing results. 


\noindent{\bf Approach.} We consider \tbase and \llama in this RQ because of their high log parsing accuracy and representative model size. 
For every system, we pre-train
the LLM using 15 {\em other} systems by following the same fine-tuning process as done before. 
For the first part of the evaluation, we apply the pre-trained \logllms directly to parse the logs of the target system (the system of which the logs are not used for pre-training). 
Then, we further fine-tune the pre-trained \logllms using 25 log samples from the target system and evaluate the accuracy of the parsed logs.  


\begin{table}\centering
\caption{Grouping (GA) and parsing (PA) accuracy of using pre-trained \logllms (i.e., {\em pt}), and \logllms that is fine-tuned based on the pre-trained \logllms (i.e., ft). }\label{tab: RQ4}
\vspace{-1em}
\scriptsize
\begin{adjustwidth}{-0.15cm}{}
\scalebox{1}{
\begin{tabular}{l|cc|cc?cc|cc}\toprule
&\multicolumn{4}{c?}{\textbf{\llama}} &\multicolumn{4}{c}{\textbf{\tbase}} \\\cline{2-9}
&GA$_{pt}$ &PA$_{pt}$ &GA$_{ft}$ &PA$_{ft}$ &GA$_{pt}$ &PA$_{pt}$ &GA$_{ft}$ &PA$_{ft}$ \\\hline
Android &0.9325 &0.7965 &0.7655 &0.6475 &0.6805 &0.6575 &0.9455 &0.9230 \\
Apache &1 &0.9940 &0.7245 &0.7300 &1 &1 &1 &1 \\
BGL &0.8250 &0.5565 &0.4415 &0.8230 &0.5840 &0.9480 &0.9575 &0.9715 \\
Hadoop &0.9595 &0.5030 &0.3270 &0.4105 &0.9560 &0.6780 &0.9955 &0.9810 \\
HDFS &0.1335 &0.2000 &0.2710 &0.7790 &0.7470 &0.4590 &1 &1 \\
HealthApp &0.6885 &0.6825 &0.6250 &0.7560 &0.7635 &0.7560 &0.8005 &0.9400 \\
HPC &0.9445 &0.8745 &0.6490 &0.8875 &0.9150 &0.9210 &0.9780 &0.9895 \\
Linux &0.5440 &0.5185 &0.1755 &0.7350 &0.5440 &0.5275 &0.4870 &0.9415 \\
Mac &0.7620 &0.4805 &0.2065 &0.5005 &0.7350 &0.4335 &0.8910 &0.7240 \\
OpenSSH &0.2280 &0.8710 &0.1660 &0.9105 &0.3900 &0.7875 &0.5020 &0.9745 \\
OpenStack &0.3740 &0.4235 &0.0385 &0.6680 &0.2670 &0.7950 &0.9890 &0.9915 \\
Proxifier &0.0010 &0 &0 &0.1730 &0.0010 &0.0005 &1 &1 \\
Spark &0.9030 &0.9010 &0.1195 &0.8755 &0.7755 &0.8980 &1 &1 \\
Thunderbird &0.6615 &0.8430 &0.0825 &0.1055 &0.9465 &0.8135 &0.6955 &0.9550 \\
Windows &0.4015 &0.5795 &0.2780 &0.8620 &0.9890 &0.9810 &1 &0.9970 \\
Zookeeper &0.8045 &0.8770 &0.7470 &0.8955 &0.9600 &0.6775 &0.9935 &0.9930 \\
\hline
Average &0.6352 &0.6313 &0.3511 &0.6724 &0.7034 &0.7083 &0.8897 &0.9613 \\
\bottomrule
\end{tabular}}
\end{adjustwidth}
 \vspace{0.5em}
\end{table}

\phead{Results.} {\bf {\em \logllms pre-trained using logs from other systems achieve considerably lower GA and PA compared to \logllms that use few-shot tuning
}} Table~\ref{tab: RQ4} shows the GA and PA of using the pre-trained \llama and \tbase. By pre-training using only the logs from 15 other systems, \llama achieves a PA and GA of 0.6352 and 0.6313, and \tbase achieves a PA and GA of 0.7034 and 0.7083. 
Compared to few-shot tuning using logs from the same system, the PA and GA for the pre-trained \logllms decrease considerably. 
Interestingly, despite having more parameters, \llama achieves lower GA and PA compared to \tbase. 
The reason may be that, compared to \llama, \tbase converges faster and is better at avoiding underfitting during fine-tuning due to its smaller parameter size. 
Nevertheless, we find that the GA and PA of the pre-trained \logllms are still comparable to that of the traditional log parsers (i.e., Drain and Logram), which further shows the potential of using LLMs for log parsing. 
{\em Our finding shows that logs from different systems may have different characteristics, so using only logs from other systems gives worse GA and PA. }

\noindent {\bf {\em Further tuning using logs from the target system shows opposite results in different \logllms. The GA and PA improved considerably in \tbase but the GA of \llama decreased by almost 55\% compared to using only the pre-trained \llama.}} 
We further fine-tune the pre-trained \logllms using 25 log samples from the target system. We find that few-shot tuning the pre-trained \logllms improved the GA and PA of \tbase to 0.8897 and 0.9613, respectively (Table~\ref{tab: RQ4}). The improvement in GA and PA is large compared to using only the pre-trained \tbase (from a GA and PA of around 0.7 to 0.8897 and 0.9613 in GA and PA, respectively). The GA and PA are also comparable to using 100 log samples from the target system to fine-tune \tbase, which has a GA and PA of 0.89 and 0.98, as shown in Table~\ref{tab:RQ1}. 
This enables us to minimize the time cost and manual effort required for labelling training data and also reduces the fine-tuning time.
However, we see an opposite result in \llama, where further tuning using 25 log samples from the target system results in worse GA (decreased from 0.6352 to 0.3511, a 55\% decrease) and similar PA (increased from 0.6313 to 0.6724, a 6\% increase). 

Some studies~\cite{nakkiran2021deep,eldan2023tinystories,chia2023instructeval} have shown that the amount of data and model parameter size required to achieve optimal performance on certain tasks are not directly correlated across models with different parameter sizes and architectures. Sometimes, having too much or too little amount of data can result in model overfitting and poor model performance. The pre-trained \llama may also need more data to fine-tune due to its larger size. Future research should explore the use of different quantities or types of log data for pre-training models based on our study in order to build more generalized, high-accuracy log parsing models that require fewer fine-tuning samples and better align with practical needs.

\rqboxc{Although pre-trained \logllms achieves worse results compared to few-shot tuning, they achieve similar results compared to the prior state-of-the-art. Further tuning the pre-trained \logllms shows opposite results in two LLMs, where the result became worse for \llama but better for \tbase. }

\section{Discussion}
\label{sec:Discussion}
\logllms achieve promising results with higher parsing accuracy than state-of-the-arts and comparable grouping accuracy with LogPPT~\cite{le2023log}. However, we also find some limitations and potential improvements in LLM-based log parsing. In this section, we summarize our observations and highlight future research directions.

\noindent{\bf {\em \logllms face challenges in parsing some specific logs. More advanced or log-tailored LLMs are needed to further improve LLM-based log parsers. }}
\logllms takes a few training samples as input and then learns how to parse logs.
However, as shown in Table~\ref{tab:RQ3}, even though \logllms achieve high accuracy (above 90\% PA), there are still some logs that were not parsed correctly. Through manual investigation, we find that \logllms face challenges in recognizing specific data types (e.g., datetime) as variables.
For instance, \llama fails to parse the log template ``\texttt{connection from <*> at <*>}'' correctly. Instead, the logs are parsed as ``\texttt{connection from <*> at Mon Jul 25 23:24:09 2005}'' without recognizing the timestamp value as the second variable.
Due to having a limited number of samples, the LLMs are not able to learn how to parse some variables. One potential solution is, similar to RQ4, to use log data from other systems to help pre-train an LLM-based log parser so that the parsers can generalize and identify more variables. The other potential solution is to use a more complex LLM with more parameters. However, future studies should consider the trade-off between more complex LLMs and higher fine-tuning and inference costs.

\noindent{\bf {\em Since more complex LLMs may not always give better results, future studies are needed to find the balance between accuracy and efficiency. }}
We find that simpler models, such as T5, can already achieve promising log parsing results and larger models may not give better results. 
There may be significant cost implications when using larger or even commercial LLMs. More importantly, as we found, larger LLMs also need more inference time to parse logs. Since logs are often large in volume, having an efficient parser is important.  
Future research should explore the right balance in model size and parsing efficiency. 




\noindent{\bf {\em Future research should explore the most effective sampling algorithms for identifying training log samples. }}
In RQ3, we find that \logllms have worse parsing results on unseen log templates and there are cases where increasing shot sizes result in worse parsing results. For example, when two very similar yet different log templates (e.g., one template has one more variable) are included in the training, the models may get confused when parsing the corresponding logs. Hence, future research should explore the optimal sampling strategy that can maximize diversity while also considering the characteristics of the logs and the corresponding templates (e.g., should certain types of logs be sampled more to distinguish the similar templates).  


\noindent{\bf {\em Fine-tuning with samples from the target system is demonstrated as the most effective way for log parsing. }}
In RQ2, we find that in-context learning shows bad performance on log parsing. Also, in-context learning is not as effective as fine-tuning due to the token limitation and efficiency concerns. 
In RQ3 and RQ4, we also find that \logllms face challenges in generalizing parsing unseen logs. 
We found that fine-tuning LLM with samples from the target system gives the best result. Future studies should consider validating the findings on more complex LLMs (e.g., LLaMA 70B) and see if such models have better generalizability on log parsing. 



\section{Threats to Validity}
\label{sec:threat}
\noindent{\bf External validity.}
Similar to prior work~\cite{le2023evaluation,le2023log,10.1145/3510003.3510101}, we train and validate \logllm using logs and log templates from public datasets that are commonly used in log-related research, However, recent research~\cite{10.1145/3510003.3510101} has indicated the dataset might contain data errors. To mitigate this potential issue, we leverage the corrected dataset~\cite{10.1145/3510003.3510101} to reduce such a threat. 
For a fair comparison, we also compare our results with the results from the state-of-the-art based on the corrected data~\cite{10.1145/3510003.3510101, le2023log}.
The log format may also affect our result, but the used datasets cover logs from various systems with different formats. Future studies are needed to evaluate LLM-based parsers on logs from other systems. 
\noindent{\bf Internal validity.}
While \logllm outperforms other state-of-the-art approaches, our primary focus in this research is on the exploration of the performance of LLMs in log parsing tasks. Our research does not cover all LLMs and training data sizes. Future studies may explore the optimal solution for LLM-based log parser, enabling further advancements in this domain. 
\noindent{\bf Construct validity.} Our approach requires pairs of logs and their log templates. Hence, the sampling process may affect the parsing result. To mitigate the issue, we apply an unsupervised data sampling algorithm that does not require any knowledge of the ground truth. Future studies are needed to explore the effect of such sampling algorithms on the parsing results. 

\section{Conclusion}
\label{sec:conclusion}

In this study, we explore the potential of leveraging LLMs for log parsing.
We propose \logllm, a generative LLM-based log parser, to overcome the limitations of existing log parsers.
\logllm leverages few-shot tuning to learn from a limited set of training logs
, which were 
sampled using a clustering sampling algorithm.
Our evaluation shows that \logllms achieve high accuracy, outperforming state-of-the-arts log parsers. We then evaluate \logllms under different pre-training settings. Our results show that, compared to in-context learning, few-shot tuning achieves higher parsing accuracy and requires less inference time.
Furthermore, our findings suggest that different \logllm models may require different numbers of training samples to achieve optimal performance. Instead of increasing the training shot sizes, future studies should investigate how training log diversity and coverage affect log parser accuracy. 
Our exploratory study leverages generative LLMs for log parsing and delivers comprehensive evaluations in various settings (architecture, shot sizes, and pre-training), which provides empirical evidence for future research.




\newpage
\balance

\bibliographystyle{ACM-Reference-Format}
\bibliography{ref}

\end{document}